\documentclass[useAMS,natbib,multicolumn,subfigure]{mn2e}
\usepackage{amsmath}
\usepackage[T1]{fontenc}
\usepackage{ae,aecompl}

\usepackage[latin1]{inputenc}  
\usepackage{graphicx,amsmath,xcolor}
\usepackage{amssymb}
\usepackage{indentfirst}
\usepackage{hyperref}
\usepackage{graphicx,pst-all}
\usepackage{tikz}
\usepackage{layout}
\usepackage{fancyhdr}

\newcommand{\kms}{\hbox{\kern 0.20em km\kern 0.20em s$^{-1}$}}
\newcommand{\cmt}{\hbox{\kern 0.20em cm$^{-3}$}}
\newcommand{\cmd}{\hbox{\kern 0.20em cm$^{-2}$}}

\def\msol{\hbox{\kern 0.20em $M_\odot$}}
\def\lsol{\hbox{\kern 0.20em $L_\odot$}}
\def\rsol{\hbox{\kern 0.20em $R_\odot$}}
\def\sr{\hbox{\kern 0.20em sr}}
\def\srmu{\hbox{\kern 0.20em sr$^{-1}$}}

\def\g{\hbox{\kern 0.20em g}}
\def\gmu{\hbox{\kern 0.20em g$^{-1}$}}
\def\kg{\hbox{\kern 0.20em kg}}
\def\pc{\hbox{\kern 0.20em pc}}

\def\mum{\hbox{\kern 0.20em $\mu$m}}
\def\mumd{\hbox{\kern 0.20em $\mu$m$^{-2}$}}
\def\cm{\hbox{\kern 0.20em cm}}
\def\m{\hbox{\kern 0.20em m}}
\def\km{\hbox{\kern 0.20em km}}
\def\nm{\hbox{\kern 0.20em nm}}

\def\s{\hbox{\kern 0.20em s}}
\def\h{\hbox{\kern 0.20em h}}
\def\sec{\hbox{\kern 0.20em sec}}
\def\min{\hbox {\kern 0.20em min}}
\def\smu{\hbox{\kern 0.20em s$^{-1}$}}
\def\smd{\hbox{\kern 0.20em s$^{-2}$}}
\def\an{\hbox{\kern 0.20em an}}
\def\anmu{\hbox{\kern 0.20em an$^{-1}$}}
\def\deg{\hbox{\kern 0.20em $^{\rm o}$}}
\def\yr{\hbox{\kern 0.20em yr}}
\def\yrmu{\hbox{\kern 0.20em yr$^{-1}$}}
\def\Myr{\hbox{\kern 0.20em Myr}}
\def\Mymu{\hbox{\kern 0.20em Myr$^{-1}$}}
\def\K{\hbox{\kern 0.20em K}}
\def\pcmu{\hbox{\kern 0.20em pc$^{-1}$}}
\def\pcmd{\hbox{\kern 0.20em pc$^{-2}$}}
\def\pcmt{\hbox{\kern 0.20em pc$^{-3}$}}
\def\kms{\hbox{\kern 0.20em km\kern 0.20em s$^{-1}$}}
\def\kmpd{\hbox{\kern 0.20em km$^{2}$}}
\def\kpc{\hbox{\kern 0.20em kpc}}
\def\cms{\hbox{\kern 0.20em cm\kern 0.20em s$^{-1}$}}
\def\erg{\hbox{\kern 0.20em erg}}
\def\ergs{\hbox{\kern 0.20em erg}}
\def\cmpd{\hbox{\kern 0.20em cm$^2$}}
\def\cmmd{\hbox{\kern 0.20em cm$^{-2}$}}
\def\cmms{\hbox{\kern 0.20em cm$^{-6}$}}
\def\cmpt{\hbox{\kern 0.20em cm$^3$}}
\def\cmmt{\hbox{\kern 0.20em cm$^{-3}$}}
\def\mpd{\hbox{\kern 0.20em m$^2$}}
\def\mmd{\hbox{\kern 0.20em m$^{-2}$}}
\def\mpt{\hbox{\kern 0.20em m$^3$}}
\def\mmt{\hbox{\kern 0.20em m$^{-3}$}}
\def\mujy{\hbox{\kern 0.20em $\mu$Jy}}
\def\mjy{\hbox{\kern 0.20em mJy}}
\def\Mj{\hbox{\kern 0.20em MJy}}
\def\jy{\hbox{\kern 0.20em Jy}}
\def\ghz{\hbox{\kern 0.20em GHz}}
\def\srmd{\hbox{\kern 0.20em sr$^{-1}$}}

\def \mum{$\mu$m}
\def\G{\hbox{\kern 0.20em G}}

\def\htwo{\hbox{H${}_2$}}
\def\h13cop{\hbox{H$^{13}$CO$^{+}$}}

\def\h2o{\hbox{H$_2$O}}

\begin{document}

\title[]{A search for Cyanopolyynes in L1157-B1}

\author[E. Mendoza et al.]{
Edgar Mendoza$^{1,2}$,
B. Lefloch$^{2,1}$,
C. Ceccarelli$^{2}$,
C. Kahane$^{2}$ ,
A. A. Jaber$^{2,3}$,
\newauthor
L. Podio $^4$,
M. Benedettini$^5$,
C. Codella$^4$,
S. Viti$^6$,
I. Jimenez-Serra$^{7}$,
\newauthor
J. Lepine$^1$,
H. M. Boechat-Roberty$^8$,
R. Bachiller$^9$
\\
$^{1}$Instituto de Astronomia, Geof\'isica e Ci\^encias Atmosf\'ericas, Universidade de S\~ao Paulo, S\~ao Paulo 05508-090, SP, Brazil\\
$^{2}$Univ. Grenoble Alpes, CNRS, IPAG, F-38000 Grenoble, France\\
$^{3}$University of AL-Muthanna, College of Science, Physics Department, AL-Muthanna, Iraq\\
$^{4}$INAF, Osservatorio Astrofisico di Arcetri, Largo Enrico Fermi 5, I-50125 Firenze, Italy\\
$^{5}$INAF, Istituto di Astrofisica e Planetologia Spaziali, via Fosso del Cavaliere 100, 00133 Roma, Italy\\
$^{6}$Department of Physics and Astronomy, University College London, Gower Street, London, WC1E 6BT, England \\
$^{7}$School of Physics and Astronomy, Queen Mary, University of London, Mile End Road, London E1 4NS, England\\
$^{8}$Observat\'orio do Valongo, Universidade Federal do Rio de Janeiro - UFRJ, Rio de Janeiro 20080-090, RJ, Brazil\\
$^{9}$IGN, Observatorio Astron\'omico Nacional, Calle Alfonso XII, 3 E-28004 Madrid, Spain\\
}

 \date{ Accepted 2018 January 5. Received 2017 November 7; in original form 2017 September 12}
\pubyear{2018}

\label{firstpage}
\pagerange{\pageref{firstpage}--\pageref{lastpage}}
\maketitle

\begin{abstract}
We present here a systematic search for cyanopolyynes in the shock region L1157-B1 and its associated protostar L1157-mm in the framework of the Large Program
"Astrochemical Surveys At IRAM" (ASAI), dedicated to chemical surveys of solar-type star forming regions with the IRAM 30m telescope. Observations of the millimeter windows between 72 and 272 GHz permitted the detection of HC$_3$N and its 13C isotopologues, and  HC$_5$N (for the first time in a protostellar shock region). In the shock, analysis of the line profiles shows that the emission arises from the outflow cavities associated with L1157-B1 and L1157-B2. Molecular abundances and excitation conditions were obtained from analysis of the Spectral Line Energy Distributions under the assumption of Local Thermodynamical Equilibrium  or using a radiative transfer code in the Large Velocity Gradient  approximation.  Towards L1157mm, the HC$_3$N emission arises from the cold envelope ($T_{rot}=10\K$) and a higher-excitation region ($T_{rot}$= $31\K$) of smaller extent around the protostar. We did not find any evidence of $^{13}$C or D fractionation enrichment towards L1157-B1.  We obtain a relative abundance ratio HC$_3$N/HC$_5$N of 3.3 in the shocked gas.
We find an increase by a factor of 30 of the HC$_3$N abundance between the envelope of L1157-mm and the shock region itself.
Altogether, these results are consistent with a scenario in which the bulk of HC$_3$N was produced by means of gas phase reactions in the passage of the shock.  This scenario is supported by the predictions of a parametric shock code coupled with the chemical model UCL\_CHEM.
\end{abstract}


\maketitle
\section{Introduction}

Cyanopolyynes (HC$_{2n+1}$N) are molecules found in a
wide range of environments in our Galaxy, from comets in our Solar
System (Crovisier et al. 2004) to molecular clouds (Takano et al
1998, 2014) and evolved stars (Cernicharo \& Gu\'elin 1996). In
star-forming regions, only relatively short chains have been reported
in the literature so far, up to HC$_{11}$N, in dense cold cores (Bell et al. 1997) and
hydrocarbon-rich protostellar envelopes of low-mass (WCCC) sources
(e.g. Sakai et al. 2008; Cordiner et al. 2012; Friesen et al. 2013).

An important property of cyanopolyynes is their stability against strong radiation field and cosmic rays (Clarke \& Ferris 1995). Recently, Jaber Al-Edhari et al. (2017) showed how the cyanopolyyne emission properties could be used to trace the history of the IRAS16293-2422 protostellar envelope.

The presence of cyanopolyynes in protostellar shock regions has received relatively little attention, since the detection of HC$_3$N by  Bachiller \& Perez-Gut\'{\i}errez (1997) in the outflow shocks of L1157, a low-mass star forming region at $\sim$250~pc (Looney et al. 2007). They estimated  an increase of the molecular abundance by about 2 orders of magnitude between the outflow shock regions L1157-B1/B2 and the position of the protostar L1157-mm. Their estimate was based on a simple LTE analysis of 3 transitions, only. No evidence for more complex cyanopolyyne (HC$_5$N) was found.
The survey of L1157-B1 in the 3mm window with the 45m telescope of the Nobeyama Radio Observatory by Yamaguchi et al. (2012) only confirmed the above results, and  no substantial advances were made in the chemistry of cyanopolyynes and the presence of longer chains.

The outflow shock region L1157-B1 and the outflow driving protostar L1157-mm have been the subject of a thorough study by our team, in particular as part of the observational Large Program
\lq\lq Astrochemical Surveys At IRAM\rq\rq \" (ASAI\footnote{http://www.oan.es/asai/}; Lefloch et al. 2017 in prep.) with the IRAM 30m telescope.
These various studies have provided us with a unique dataset, which offers the possibility of investigating  the physical and chemical structure of a typical protostellar shock (Lefloch et al. 2010, 2012, 2016, 2017; Benedettini et al. 2012, 2013; Codella et al. 2010, 2012, 2015, 2017; Busquet et al. 2014; Podio et al. 2014, 2016, 2017).

In this work, we present an observational study on the chemistry of  cyanopolyynes,  in a shock region, as well as the impact of protostellar shock on the chemical conditions in the ambient cloud. The paper is organized as follows. In Sect.~2, we summarize the main observational properties of the shock region L1157-B1. The observations are described in Sect.~3. In Sect.~4, we present the molecular transitions of cyanpolyynes detected in the ASAI line survey. The properties of the shock, the origin of the emission and the shock physical conditions (density, temperature, abundance) are discussed in Section~5. The cyanopolyyne emission towards the protostar is presented in Section~6. We discuss our results in Section~7. Finally, we present our conclusions in Section~8.

\section{The L1157-B1 shock region}
\label{source}

\begin{table*}
\centering{
\caption{Outflow shock components and their physical conditions in the L1157-B1 shock region, from CO and CS emission. The \htwo\ column densities were obtained from the CO column densities derived by
Lefloch et al. (2012), adopting a CO/$\htwo$ abundance ratio of $10^{-4}$ (see also G\'omez-Ruiz et al. 2015).}
\begin{tabular}{lrrrrrrr}\hline
       & Vel. range & $v_0$    & T       & n($\htwo$)      & N(\htwo) & Size           & Origin \\
       &   ($\kms$) & ($\kms$) & ($\K$)  & ($\cmmt$)       & ($\cmmd$)&  ($\arcsec$)   & \\ \hline
$g_1$  &  -40; 0    & 12       & 210     & $\geq 1.0(6)$   & 0.9(20)  & 7--10          & jet impact region\\
$g_2$  &  -20; 0    & 4        &  64     & $\sim$  1.0(6)  & 0.9(21)  & 20             & B1 outflow cavity walls \\
$g_3$  &  -8; 0     & 2        & 23      & (0.5--2)(5)     & 1.0(21)  & Extended       & B2 outflow cavity walls \\ \hline
\end{tabular}
}
\end{table*}

\begin{figure}
\centering{
\caption{The L1157-B1 bow-shock region as traced by the HC$_3$N(16--15) emission, observed with the
PdB interferometer (from Benedettini et al. 2013). The location of the individual shocked clumps B1a-b-c and B0e
indicated (Benedettini et al. 2013). The arrow points to the direction of the L1157-mm protostar. The
first contour and steps correspond to 5$\sigma$ (170 mJy~beam$^{-1}$~km~s$^{-1}$) and 3$\sigma$,
 respectively. }
\label{map-hc3n}
\includegraphics[width=0.9\columnwidth]{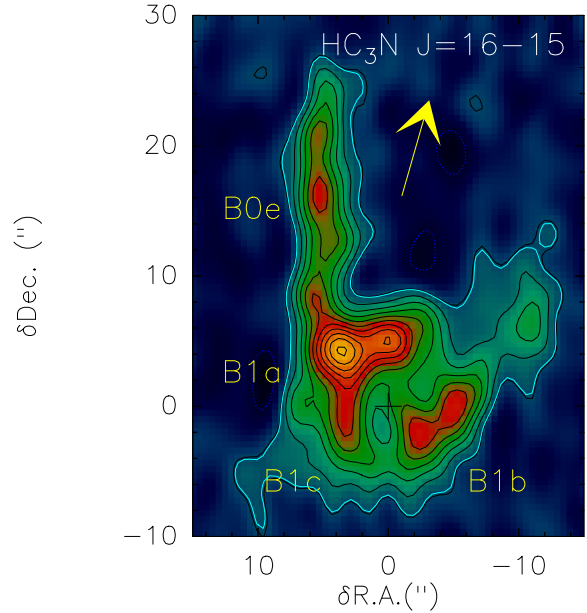}
}
\end{figure}

The L1157 bipolar and chemically rich molecular outflow (Bachiller et al. 2001) is swept
up by an episodic and precessing jet (Gueth et al. 1996, 1998; Tafalla et al. 2015; Podio et al. 2016)
driven by the L1157-mm Class 0 protostar.
The blue-shifted southern lobe mainly consists of two main cavities with different kinematical ages:
B1 ($\approx$ 1100 yr), and the older and more extended B2 (1800 yr).
The bright bow shock at the B1 cavity  is located at $\sim$ 69$\arcsec$ (0.1 pc) from
the protostar (see Fig. 1), and various molecular tracers indicate warm shocked gas enriched by the injection of dust mantle and core products, such as
CH$_3$OH, NH$_3$, H$_2$CO, NH$_2$CHO (see e.g. Tafalla \& Bachiller 1995; Benedettini et al. 2012; Codella et al. 2010, 2017; Mendoza et al. 2014; Lefloch et al. 2017; and references therein).
High-angular resolution observations of these tracers with the Plateau de Bure Interferometer (PdBI) reveal the presence of a few high-velocity clumps (B1a-b-c and B0e) whereas the lower velocity material traces the expansion of the cavity excavated by the shock (see e.g. Benedettini et al. 2013).

Previous analysis of the CO and CS gas kinematics, as observed with the IRAM 30m telescope
(Lefloch et al. 2012; G\'omez-Ruiz et al. 2015), has led to identify
three physically distinct components coexisting in L1157-B1, labelled as follows and summarized in Table~1:
\begin{itemize}
\item~$g_1$: the region of $\sim$7$\arcsec$--10$\arcsec$ at $T_{\rm kin}$ $\sim$ 210 K
and $n_{\rm H_2}$ $\geq$ 10$^6$ cm${-3}$, associated with
a dissociative J-shock due to the youngest impact of the jet against the B1 cavity walls.
\item~$g_2$: the outflow cavity walls ($\sim$ 20$\arcsec$, 60--80 K, 10$^5$--10$^6$ cm$^{-3}$) associated with the B1 ejection.
\item~$g_3$: the older outflow cavity walls (extended, 20 K, $\simeq$ 10$^5$ cm$^{-3}$)
associated with the B2 ejection.
\end{itemize}
Lefloch et al. (2012) and G\'omez-Ruiz et al. (2015) showed that each component is characterized by
homogeneous excitation conditions independent of the velocity range (see Table 1). The intensity-velocity distribution
of each component is well described by an exponential law $T_b(v) \propto exp(v/v_0)$, where $v_0$ is
almost independent of the molecular transition considered ($v_{\rm 0}$ = 12.5, 4.4, and 2.5 km s$^{-1}$ for
$g_1$, $g_2$, and $g_3$, respectively).

\begin{figure*}
\begin{center}
\includegraphics[width=1.8\columnwidth]{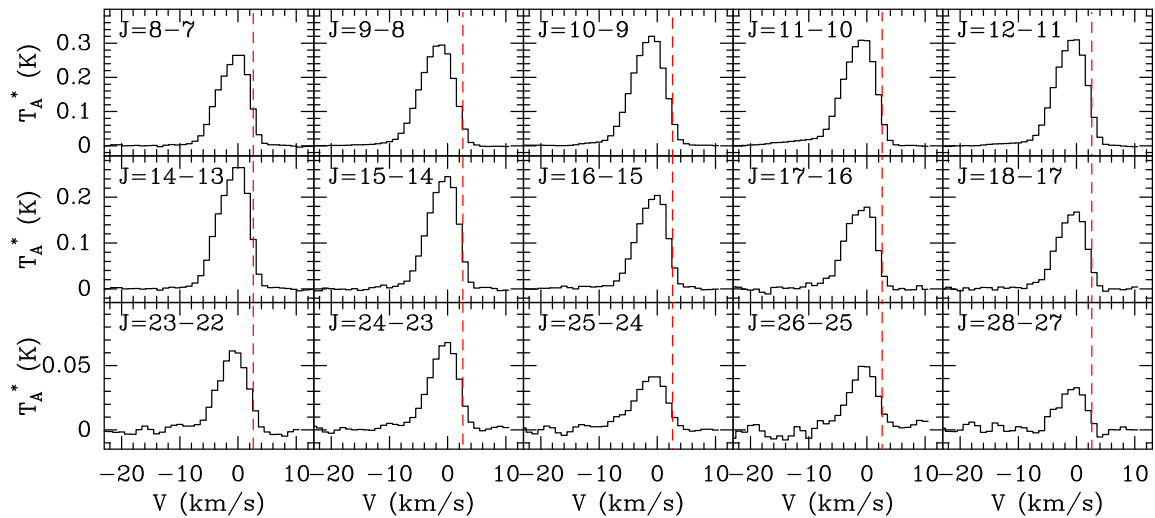}
\caption{Montage of some HC$_3$N lines detected towards L1157-B1. The red line (dashed) draws the velocity of the cloud
($v_{lsr}=+2.6\kms$).
}
\label{panel-hc3n}
\end{center}
\end{figure*}

\section{Observations}

Observations of the IRAM Large Program ASAI were carried out during several runs between September 2012 and  March 2015. The nominal position observed are  $\alpha_{J 2000} =$ 20$^{\text h}$ 39$^{\text m}$ 10.$^{\text s}$2, $\delta_{J 2000} =$ +68$^{\circ}$ 01$^{\prime}$ 10$^{\prime\prime}$ and
$\alpha_{J 2000} =$ 20$^{\text h}$ 39$^{\text m}$ 06.$^{\text s}$3, $\delta_{J 2000} =$ +68$^{\circ}$ 02$^{\prime}$ 15.8$^{\prime\prime}$ for L1157-B1 and L1157-mm, respectively.

Data were collected using the broad-band EMIR (Eight MIxer Receiver) receivers at 3~mm, 2~mm, 1.3~mm,  whose interval frequencies are 72 -- 116~GHz, 128 --173~GHz, 200 -- 272~GHz, respectively. Both the Fast Fourier Transform Spectrometers and the WILMA backends were connected to the EMIR receivers providing a spectral resolution of $\sim$~200~kHz and 2~MHz, respectively. The FTS spectral resolution was degraded to a final velocity resolution of $\approx 1\kms$. In order to ensure a flat baseline across the spectral bandwith observed, the observations were carried out in Wobbler Switching Mode, with a throw of~$3^{\prime}$.

For the ASAI data, the reduction was performed using the GILDAS/CLASS software\footnote{https://www.iram.fr/IRAMFR/GILDAS/}.
The calibration uncertainties are typically 10, 15, and 20\% at 3~mm, 2~mm and 1.3~mm,  respectively. The line intensities  are expressed in units of antenna temperature corrected for atmospheric
attenuation and rearward losses ($T_A^{\ast}$). For subsequent analysis, fluxes are expressed in main beam
temperature units ($T_{mb}$). The telescope and receiver parameters (beam efficiency, Beff; forward efficiency,
Feff; Half Power beam Width, HPBW) were taken from the IRAM webpage\footnote{http://www.iram.es/IRAMES/mainWiki/Iram30mEfficiencies}.

\begin{figure}
\centering{
\includegraphics[width=0.85\columnwidth]{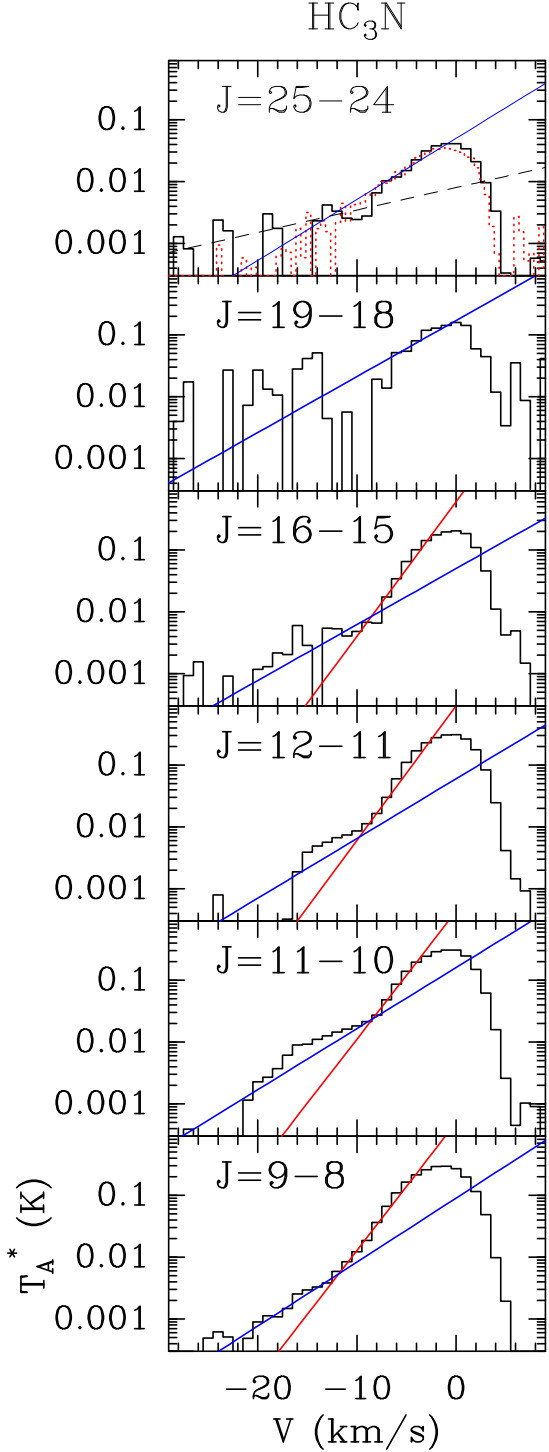}
\caption{Montage of some HC$_3$N spectra detected towards L1157-B1. We have superimposed the fit obtained by a linear combination of two exponential functions $g_2 \propto \exp(-v/4.0)$ (blue) and $g_3 \propto \exp(-v/2.0)$ (red). In the top panel ($J=25-24$), a fit of the $g_1$ component ($\propto \exp(-v/12)$) is superimposed in black and the CS $J$=7--6 line (a proxy for the $g_2$ component) is superimposed in  dotted red.
\label{profiles}}}
\end{figure}

\section{Cyanopolyyne emission in L1157-B1}

The ASAI survey permitted the identification of several  rotational transitions of HC$_3$N and HC$_5$N. The sensitivity of the survey permitted the detection of a few transitions of the $^{13}$C isotopologues of HC$_3$N.  We failed to detect larger cyanopolyynes as a search for HC$_7$N yielded only negative results. In this section, we describe the spectral signatures of HC$_3$N and HC$_5$N.

\subsection{HC$_3$N}

We detected all the HC$_3$N rotational transitions falling
in the ASAI bands, from $J$=8--7 ($E_{up}$= $16\K$) up to $J$=29--28 ($E_{up}$= $189.9\K$). The spectra  are displayed in  Fig.~2. The line spectroscopic and observational properties are summarized in Table~2.

The bulk of the emission peaks at $-1\kms$, with a linewidth (FWHM) ranging between 5 and $6\kms$ (see Table~2).
In the transitions with $J_{up}\leq 16$,  emission is detected up to $v_{lsr}$= $-20 \kms$.
The intensity-velocity distribution can be well fit adopting a relation of the type
$T\propto exp(v/v_0)$ with  $v_0$= $+4\kms$,  independently of the rotational number of the transition (Fig.~\ref{profiles}).
This spectral signature has been observed in all the transitions, up to $J$=25--24. In the lower $J$ transitions, from
$J$=8--7 to $J$=19--18, a second emission component with a lower velocity range, between 0 and $-8\kms$, is detected. It can be fit by a second exponential component with an exponent $v_0$= $+2\kms$, which, again, does not vary with $J_{up}$ (Fig.~\ref{profiles}).

This confirms that at least the high-$J$ HC$_3$N line emission arises from  component $g_2$, i.e. the outflow cavity associated with L1157-B1.  In the top right panel of Fig.~\ref{profiles}, we have superimposed the CS $J$=7--6 spectrum (dashed red) on the HC$_3$N $J$=25--24 line profile, applying a scaling factor so to match the peak intensity.  We find an excellent match between both line profiles over the full emission range. This indicates that the high-$J$  ($J\geq 25$) HC$_3$N line emission arises from component $g_2$. This point is further addressed in Sect.~5.

This profile decomposition is justified by the fact that the HC$_3$N
line emission is optically thin, except at velocities close to that of the ambient cloud for the low-$J$ transitions. We note however that the frequency of the transitions $J_{up}$= 8--12 is low enough that the telescope beam encompasses the shock region B0 (Fig.~1), which is associated with another ejection (Bachiller et al. 2001), so that the emission collected, and the line profiles,  could be contaminated by this shock region. This point is addressed further below in Sect.~5.3.

\begin{table*}
\centering{
  \caption{L1157-B1. Spectroscopic  and observational parameters of the lines from  HC$_3$N, its isotopologues, and HC$_5$N identified towards the protostellar shock L1157-B1. We give in bracket the statistical uncertainties on the observational parameters, as derived from a gauss fit to the profiles.}
\begin{tabular}{lrrcccrrrr}
\hline
Transition  & Frequency & $E_u$ & $A_{ul}$ & HPBW & $\eta_{mb}$  & $\int T_{mb}dv$  & FWHM & $V_{lsr}$ \\
$J\rightarrow J-1$ & MHz &  K &  s$^{-1}$  & $^{\prime\prime}$ &  &   mK km s$^{-1}$  & km s$^{-1}$ & km s$^{-1}$ \\
\hline
HC$_3$N & & & & & & & &   \\
8 -- 7&	72783.822   & 15.7&	2.94(-5)& 33.8&	0.86&	2044(3)& 5.9(0.1)& -0.7(0.1)	\\
9 -- 8&	81881.468   & 19.7&	4.21(-5)& 30.0&	0.85&	2556(4)& 6.2(0.2)& -1.7(0.1)	\\
10 -- 9& 90979.023  & 24.0&	5.81(-5)& 27.0&	0.84&	2674(4)& 6.1(0.2)& -1.4(0.1)	\\
11 -- 10& 100076.392& 28.8&	7.77(-5)& 24.6&	0.84&	2729(7)& 6.2(0.1)& -1.3(0.1)	\\
12 -- 11& 109173.634& 34.1&	1.01(-4)& 22.5&	0.83&	2560(7)& 5.9(0.1)& -1.0(0.1)	\\
14 -- 13& 127367.666& 45.9&	1.62(-4)& 19.3&	0.81&	1790(8)& 5.8(0.1)& -0.9(0.1) \\
15 -- 14& 136464.411& 52.4&	1.99(-4)& 18.0&	0.81&	1951(9)& 5.8(0.1)& -0.8(0.1) \\
16 -- 15& 145560.960& 59.4&	2.42(-4)& 16.9&	0.79&	1659(9)& 5.8(0.1)& -0.9(0.1)	\\
17 -- 16& 154657.284& 66.8&	2.91(-4)& 15.9&	0.77&  1516(19)& 5.9(0.1)& -1.1(0.1)	\\
18 -- 17& 163753.389& 74.7&	3.46(-4)& 15.0&	0.77&  1370(12)& 5.6(0.1)& -0.8(0.1)\\
19 -- 18& 172849.301& 83.0&	4.08(-4)& 14.2&	0.75&  1248(92)& 5.9(0.3)& -1.1(0.2)	\\
23 -- 22& 209230.234&120.5& 7.26(-4)& 11.8&	0.67&	 543(9)& 5.7(0.2)& -0.8(0.1)	\\
24 -- 23& 218324.723&131.0& 8.26(-4)& 11.3&	0.65&	611(10)& 5.3(0.2)& -0.6(0.1)	\\
25 -- 24& 227418.905&141.9& 9.35(-4)& 10.8&	0.64&	 430(8)& 6.1(0.3)& -1.1(0.1)	\\
26 -- 25& 236512.789&153.2& 1.05(-3)& 10.4&	0.63&	451(14)& 5.3(0.4)& -0.6(0.1)	\\
27 -- 26& 245606.320&165.0& 1.18(-3)& 10.0&	0.62&	273(10)& 7.1(0.4)& -0.9(0.2)	\\
28 -- 27& 254699.500&177.3& 1.32(-3)&  9.7&	0.60&	289(12)& 4.9(0.4)& -0.8(0.2)	\\
29 -- 28& 263792.308&189.9& 1.46(-3)&  9.3&	0.60&	187(16)& 5.0(0.8)& -0.5(0.3)	\\
\hline
H$^{13}$CCCN&	   &	   & 	     &	  &	    &			        &				&		\\
10 -- 9 & 88166.832& 23.3& 5.29(-5)&27.9& 0.85&	40.7(4.5)& 7.1(1.4)&-1.8(0.5)\\
11 -- 10& 96983.001& 27.9& 7.07(-5)&25.4& 0.85&	76.0(4.8)& 7.2(1.0)&-0.8(0.4)\\
12 -- 11&105799.113& 33.0& 9.21(-5)&23.3& 0.84& 48.1(3.9)& 5.2(0.8)&-2.3(0.3)\\
\hline
HC$^{13}$CCN&	   &     &		   &    &	  &			        &				&		 \\
9 -- 8   &81534.111& 19.6& 4.16(-5)&30.2& 0.86&	55.8(2.7)& 6.1(0.9)& -2.7(0.3)\\
10 -- 9 & 90593.059& 23.9& 5.74(-5)&27.2& 0.85&	43.1(3.7)& 6.5(1.5)& -3.8(0.5) \\
11 -- 10& 99651.849& 28.7& 7.67(-5)&24.7& 0.84&	43.6(5.2)& 5.0(1.7)& -0.4(0.7) \\
12 -- 11&108710.532& 33.9& 1.00(-4)&22.6& 0.83& 31.3(5.0)& 6.1(2.7)& -0.7(0.9)\\
\hline
HCC$^{13}$CN&	   &	 &		   &    &	  &                &               &           \\
9 -- 8  & 81541.981& 19.6& 4.16(-5)&30.2& 0.86& 25.7(3.1) & 4.5(1.4)& -0.7(0.6) \\
10 -- 9 & 90601.777& 23.9& 5.74(-5)&27.2& 0.85&	38.1(4.2) & 4.8(1.6)& -1.8(0.6) \\
11 -- 10& 99661.467& 28.7& 7.67(-5)&24.7& 0.84&	33.3(5.3) & 4.5(2.0)& -1.1(0.8) \\
12 -- 11&108720.999& 33.9& 1.00(-4)&22.6& 0.83&	35.5(3.4) & 4.0(0.7)& -0.4(0.3)\\
\hline
HC$_5$N	&		&		&		&		&		&			&				&		 \\
30 -- 29&	79876.710&	59.4& 5.47(-5)& 30.8& 0.86&	39.1(3.4)&	3.5(0.8)&	-1.5(0.3)	\\
31 -- 30&	82539.039&	63.4& 6.04(-5)& 29.8& 0.85&	22.5(1.7)&	3.7(0.8)&	-1.7(0.3)	\\
32 -- 31&	85201.346&	67.5& 6.64(-5)& 28.9& 0.85&	34.1(1.7)&  5.0(0.8)&	-1.9(0.3)	\\
33 -- 32&	87863.630&	71.7& 7.29(-5)& 28.0& 0.85&	47.1(1.5)&  7.4(1.0)&	-1.8(0.4)	\\
34 -- 33&	90525.889&	76.0& 7.98(-5)& 27.2& 0.85&	37.1(1.7)&  6.3(1.2)&	-1.5(0.5)	\\
35 -- 34&	93188.123&	80.5& 8.71(-5)& 26.4& 0.85&	30.8(2.6)&	5.4(2.3)&	-0.6(0.7)	\\
36 -- 35&	95850.335&	85.1& 9.48(-5)& 25.7& 0.85&	36.0(2.4)&	5.7(1.4)&	-0.4(0.6)	\\
37 -- 36&	98512.524&	89.8& 1.03(-4)& 25.0& 0.84&	27.9(1.2)&	3.5(0.8)&	 1.8(0.6)	\\
38 -- 37&	101174.677&	94.7& 1.12(-4)& 24.3& 0.84& 38.2(2.0)&	7.6(2.1)&	-2.2(0.9)	\\
39 -- 38&	103836.817&	99.7& 1.21(-4)& 23.7& 0.84&	31.1(2.0)&	6.0(1.6)&	-2.7(0.6)	\\
40 -- 39&	106498.910&	104.8&1.30(-4)& 23.1& 0.84&	39.2(2.2)&	5.0(1.0)&	 0.2(0.4)	\\
41 -- 40&	109160.973&	110.0&1.40(-4)& 22.5& 0.83& 28.2(2.2)&	5.0(2.1)&	-2.3(0.8) \\
42 -- 41&	111823.024&	115.4&1.51(-4)& 22.0& 0.83&	45.0(3.4)&	4.4(1.0)&	 0.2(0.4)\\
43 -- 42&	114485.033&	120.9&1.62(-4)& 21.5& 0.83&	23.0(3.9)&	2.9(1.1)&	-0.8(0.5)	\\
\hline
\end{tabular}
}
\\
%
\end{table*}

\begin{figure}
  \caption{Montage of the detected transitions from the rare isotopologues H$^{13}$CCCN, HC$^{13}$CCN and HCC$^{13}$CN.   }
  \label{isotopologues}
  \centering{\includegraphics[width=\columnwidth,keepaspectratio]{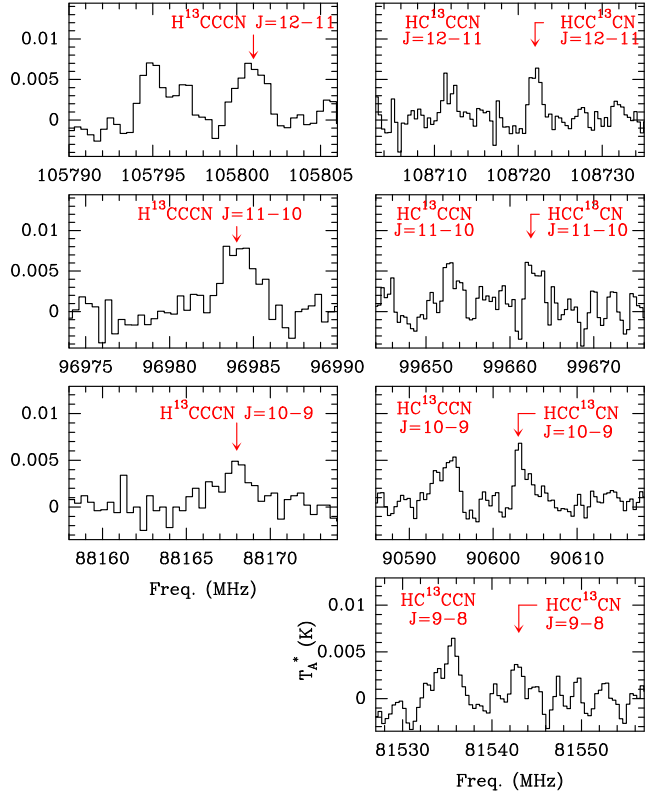}
  }
\end{figure}

The search for the HC$_3$N isotopologues yielded positive results in the band at 3~mm. We detected the transitions
H$^{13}$CCCN $J$=10--9, 11--10 and 12--11; HC$^{13}$CCN $J$=9--8, 10--9, 11--10, 12--11; and HCC$^{13}$CN  $J$=9--8, 10--9, 11--10 and 12--11, whose line profiles are displayed in Fig.~4.
Intensity of all the transitions is weak, typically 5 mK ($T_{A}^{*}$), so that the line parameters determination suffers large uncertainties. Their properties are summarized in Table~2.

We searched for emission lines from the deuterated isotopologue of HC$_3$N, following the previous work by Codella et al. (2012) on the fossile deuteration in  L1157-B1. We failed to detect any of the transitions falling in the 3mm ASAI band, down to a $3\sigma$ level of 4~mK ($T_A^{*}$).

\subsection{HC$_5$N}
\begin{figure*}
\centering{
\includegraphics[width=1.8\columnwidth]{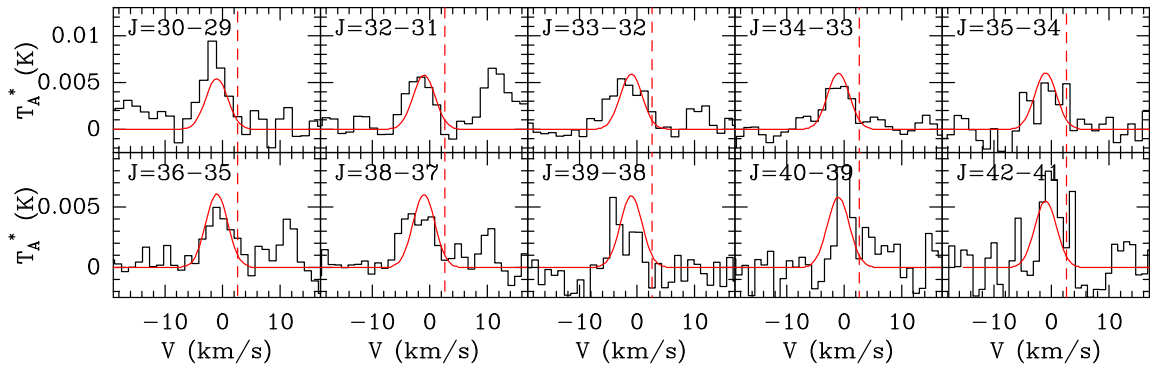}
\caption{Montage of some HC$_5$N transitions detected towards L1157-B1. The best fit obtained from  a simple rotational diagram analysis is shown in red, adopting a linewidth $\Delta v$= $5\kms$. The ambient cloud velocity ($v_{lsr}$= $+2.6\kms$) is marked by the dash-dotted line.}}
\end{figure*}

Only transitions in the 3mm band, between 80 and 114 GHz, could be detected.
All the transitions from $J$=30--29 ($E_{up}$= $59\K$) to $J$=43--42 ($E_{up}$= $121\K$) were detected down to the $3\sigma$ level (Table~2).
Line intensities are rather weak, typically 5--10mK, a factor $\sim 20$ less than HC$_3$N transitions of similar $E_{up}$.
We show a montage of some HC$_5$N lines detected towards L1157-B1 in Fig.~5. The typical line width (FWHM) and the emission peak velocity are $\simeq 4.5\kms$ and $-1\kms$, respectively, hence in good agreement with
the parameters of the HC$_3$N lines.  The analysis of the excitation conditions (see Sect.~5.3 below) leads us to conclude that the emission of HC$_5$N is dominated by $g_2$.

\section{Physical conditions and abundances in L1157-B1}
\label{physical}

The physical conditions and the column densities of HC$_3$N and HC$_5$N were
obtained from the analysis of their Spectral Line Energy Distribution
(SLED) via a Large-Velocity Gradient (LVG) modelling and under the assumption of
Local Thermodynamic Equilibrium (LTE), respectively. We discuss the
two cases separately.

\subsection{HC$_3$N}
\begin{figure}
  \centering{
    \includegraphics[width=0.85\columnwidth,keepaspectratio]{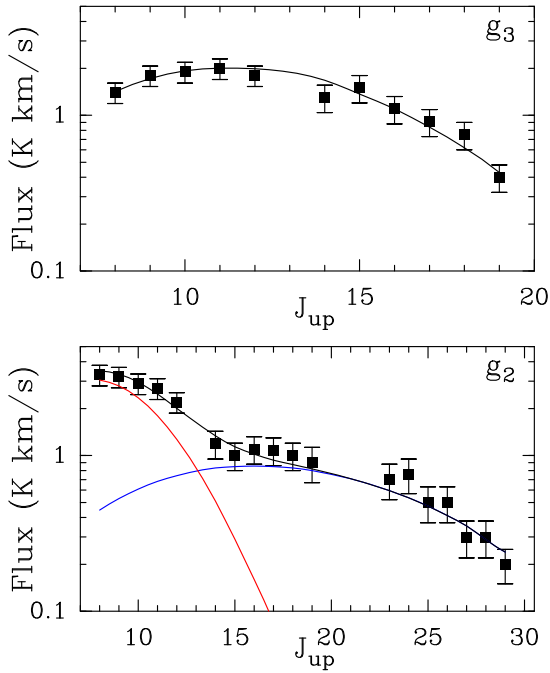}
    \caption{Modelling of the HC$_3$N Spectral Line Energy
      Distribution (SLED) with a simple LVG slab model. The black
      squares represent the measured velocity-integrated fluxes for
      components $g_2$ (bottom) and $g_3$ (top). They are expressed in
      units of main-beam brightness temperatures ($T_{mb}$).  ({\em
        Top panel:}) The solid curve shows the LVG predictions for
      component $g_3$  assuming one single gas component with a
      column density $N$(HC$_3$N)=2.5$\times 10^{13}\cmmd$, a
      temperature of 20 K and a density $n(\htwo)$
      $2\times 10^6\cmmt$. ({\em Bottom panel:}) The solid curves show
      the LVG predictions for component $g_2$, obtained assuming two components
      at $20\K$ (red), with density n($\htwo$)= $6\times 10^4\cmmt$, N(HC$_3$N)= $2\times 10^{13}\cmmd$
      and at $60\K$ (blue), with density n($\htwo$)= $4\times 10^6\cmmt$, N(HC$_3$N)= $4\times 10^{12}\cmmd$,  respectively.}
      \label{fit-hc3n-lvg}}
\end{figure}

As discussed in the previous section, the HC$_3$N line emission can be
decomposed in two components, $g_2$ and $g_3$, which probe the outflow
cavity walls of B1 and B2, respectively.  Figure \ref{fit-hc3n-lvg}
shows the Spectral Line Energy Distribution (SLED) of the two
components. While the $g_3$ SLED peaks at J$\sim$11, the $g_2$ SLED
clearly shows two different peaks, at J$\leq$8 and J$\sim$20,
suggesting the presence of a cold and a warm physical component.

Before modelling the SLED of each component, we first estimated the
brightness temperature of the molecular lines.  Our previous studies
of CO and CS showed that the component $g_3$ is extended enough that
the brightness temperature of the lines can be approximated by the
main-beam brightness temperature $T_{mb}$.

As the source size of component $g_2$ is not precisely known, we have chosen
to perform the SLED modeling at a common angular resolution of $8.5\arcsec$
(the HPBW at the frequency of the J=32-31 line, the highest detected transition
of HC$_3$N in L1157-B1), for all the transitions. It means that the effective column
density derived for the $g_2$ component scales as the effective filling factor in the
$8.5\arcsec$  beam. To do so, we followed the methodology presented in Lefloch et al. (2012) and
Gomez-Ruiz et al. (2015). The $g_2$ and $g_3$ velocity integrated fluxes used in our
SLED analysis are reported in Table 3.

Taking into account the uncertainties in the
spectral decomposition of component $g_2$ and $g_3$, we estimate the
total uncertainties to be 15\%, 20\%, 25\% at 3mm, 2mm and 1.3mm,
respectively.

To model the SLED emission, we used the LVG code by Ceccarelli et
al. (2002) and used the collisional coefficients with para and ortho
H$_2$ computed by Faure et al. (2016). For the H$_2$ ortho-to-para
ratio we assumed the fixed value of 0.5, as suggested by previous
observations.
We ran a large grid of models with density between $10^4\cmmt$ and
$10^8\cmmt$, temperature between $15\K$ and $120\K$ and the HC$_3$N
column density between $2\times 10^{12}$ and $3\times 10^{14}\cmmd$.
We adopted $\Delta V$= $5.5\kms$ and let the filling factor
to be a free parameter in the case of the $g_2$ component.
We then compared the LVG predictions and the observations and used the
standard minimum reduced $\chi^2$ criterium to constrain the four
parameters: density, temperature, column density and size (in $g_2$
only).
\begin{table}
  \centering{
    \caption{HC$_3$N velocity-integrated intensities  of the
      components $g_2$ and $g_3$ used in the radiative transfer
      calculations. Fluxes are expressed in units of brightness
      temperature in a beam of~$8.5\arcsec$ for $g_2$ and in main-beam
      brightness temperature for $g_3$, respectively. }
    \label{tab-hc3-lvg-flux}
    \begin{tabular}{crccc}
      \hline
Transition  & Frequency     & ff ($g_2$)& Flux ($g_2$)&  Flux ($g_3$)   \\
            & (MHz)         &           & K $\kms$    &  K $\kms$  \\  \hline
8 -- 7	    &	72783.822	&	0.184	&	3.34	  &	1.43	\\
9 -- 8	    &	81881.468	&	0.227	&	3.23	  &	1.83	\\
10 -- 9	    &	90979.023	&	0.268	&	2.91	  &	1.89	\\
11 -- 10	&	100076.392	&	0.310	&	2.52	  &	1.95	\\
12 -- 11	&	109173.634	&	0.354	&	2.24	  &	1.77	\\
14 -- 13	&	127367.666	&	0.437	&	1.02	  &	1.34	\\
15 -- 14	&	136464.411	&	0.482	&	0.97	  &	1.49	\\
16 -- 15	&	145560.960	&	0.522	&	1.13	  &	1.07	\\
17 -- 16	&	154657.284	&	0.563	&	1.29	  &	0.91	\\
18 -- 17	&	163753.389	&	0.603	&	1.01	  &	0.76	\\
19 -- 18	&	172849.301	&	0.642	&	0.90      &	0.40	\\
23 -- 22    &   209230.234  &   0.775   &   0.70       & 0.0  \\
24 -- 23    &   218324.723  &   0.807   &   0.76      & 0.0  \\
25 -- 24    &   227418.905  &   0.839   &   0.51      & 0.0  \\
26 -- 25    &   236512.789  &   0.866   &   0.52      & 0.0  \\
27 -- 26    &   245606.320  &   0.892   &   0.30       & 0.0  \\
28 -- 27    &   254699.500  &   0.914   &   0.32      & 0.0  \\
29 -- 28    &   263792.308  &   0.942   &   0.20       & 0.0  \\
      \hline
    \end{tabular}
  }
\end{table}
%
%
The best fitting LVG solutions are shown in Fig.~\ref{fit-hc3n-lvg}.

The $g_3$ component SLED is well reproduced by an HC$_3$N column
density $N$(HC$_3$N)= $2.5\times 10^{13}\cmmd$, a temperature of
20 K and a density $n(\htwo)$ $\simeq 2\times 10^6\cmmt$.
Adopting an \htwo\ column density of $10^{21}\cmmd$ for $g_3$ (Table~1; Lefloch et al. 2012), we find a
relative abundance X(HC$_3$N)= $2.5\times 10^{-8}$.

For the $g_2$ component, our modelling of the high-excitation lines ($J_{up}$=14--29)
shows that the best fit is obtained for N(HC$_3$N)=$4\times 10^{12}\cmmd$, $T_{kin}$= $60\K$ and
$n(\htwo)$=$4\times 10^6\cmmt$, and a filling factor of 0.82, corresponding to a source size of $18\arcsec$. This simple model fails however to reproduce the low excitation range
 $J_{up}$=8--12 of the SLED (see Fig.~\ref{fit-hc3n-lvg}). A second gas
component with a lower density $n(\htwo)$= $6\times 10^4\cmmt$ and
lower temperature T=$20\K$, is needed to account for the total
observed flux. We estimate for this component a column density
N(HC$_3$N)=$2\times 10^{13}\cmmd$, and a size of $10\arcsec$. These
physical conditions differ from those of $g_2$ and $g_3$ (see also
Sect.~2). As pointed out in Sect.~4.1, in the frequency range
corresponding to the $J$=8--12 transitions, the beam size is large
enough ($\geq 23\arcsec$) that it encompasses the shock region B0,
which thereby contributes to the collected emission. Conversely, at
frequencies higher than that of the $J$=15--14 line, the beam size
(HPBW) is about $\leq 18\arcsec$, and the telescope beam misses the B0
emission. This is probably the cause of the observed excess in the low
J emission lines. We note that the density of the high-velocity
bullets reported by Benedettini et al. (2013) is lower by a factor of
a few in the B0 region, as compared to the B1 region; it is in
agreement with the values obtained in the present analysis. Also, Lefloch et al. (2012)
showed that the $g_2$ spectral signature was actually detected along the entire
walls of the B1 cavity, between the protostar L1157-mm and the apex where bright molecular
emission is detected.
Proceeding as above for component $g_3$, we obtain an abundance of $4\times 10^{-9}\cmmd$
for HC$_3$N in component $g_2$.

To summarize,  the HC$_3$N  emission arises from two physical components, $g_3$ and $g_2$ respectively, with an abundance relative to \htwo\ of $2.5\times 10^{-8}$ and $4\times 10^{-9}$, respectively. These results are consistent with the previous work by Bachiller \& P\'erez-Gut\'{\i}errez (1997). These authors  estimated an abundance relative to \htwo\ of $1.0\times 10^{-8}$, assuming one single gas component, from the detection of the three  HC$_3$N $J$=10--9, 15--14, 14--23 lines in the shock. Interestingly, the abundance of HC$_3$N towards $g_2$ appears lower than towards $g_3$, as low as a factor of 6.

%
%

\subsection{The rare isotopologues of HC$_3$N}
\subsubsection{H$^{13}$CCCN, HC$^{13}$CCN, and HCC$^{13}$CN}
Only the transitions from H$^{13}$CCCN, HC$^{13}$CCN, and HCC$^{13}$CN with $E_{up}$ between 19 and $34\K$ were detected.   The low number (3--4) of detected transitions for each isotopologue prevents us from carrying out a detailed analysis of the emission, similar to the approach adopted for HC$_3$N.  In addition, our analysis of the HC$_3$N emission shows that in this range of $E_{up}$, the emission is dominated by $g_3$ and most probably contaminated by emission from the shock region B0, so that even a rotational diagram analysis would be somewhat not significant.

We have just simply assumed that the emission of the $^{13}$C-bearing isotopes and of the main isotope in the J=9--8 to J=12--11 lines originate from the same gas so that the corresponding line fluxes ratios provide a reliable measurement of the molecular abundance ratios if the main isotopologue is optically thin.
The mean flux ratios are $52\pm 15$, $63\pm 20$, $81\pm 10$, for  HC$^{13}$CCN, HC$^{13}$CCN and HCC$^{13}$CN, respectively.  These results do not show any significant difference between the three $^{13}$C-bearing isotopes that could indicate a differential $^{13}$C fractionation between the three HC$_3$N isotopologues. These molecular abundance ratios are also consistent with a standard value for the elemental abundance ratio $^{12}$C/$^{13}$C of 70, which is consistent with the $^{12}$C isotopologue being optically thin.

\subsubsection{DC$_3$N}

None of the DC$_3$N lines falling in the 3mm band were detected, down to an rms of 1.4~mK. We obtained an upper limit on the column density of DC$_3$N in the LTE approximation. We adopted the value $T_{rot}$= $25\K$, obtained for HC$_3$N in the same range of $E_{up}$ (18--$69\K$). Since the HC$_3$N emission is dominated by component $g_3$ in this range of $E_{up}$, we assumed DC$_3$N to arise from the same region, and to be extended. We obtain as upper limit N(DC$_3$N)= $5\times 10^{10}\cmmd$, from which we derive an upper limit on the abundance [DC$_3$N]= $5\times 10^{-11}$ and the deuterium fractionation ratio D/H= 0.002.

\subsection{HC$_5$N}
\begin{figure}
  \centering
  \includegraphics[width=0.85\columnwidth,keepaspectratio]{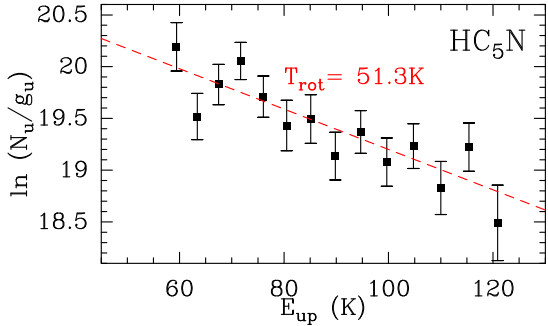}
  \caption{Rotational diagram of HC$_5$N. The emission was corrected for the coupling between the source and the telescope beam assuming a typical size of $18\arcsec$.}
    \label{rotational}
\end{figure}

The physical conditions of HC$_5$N were obtained from a simple
rotational diagram analysis, as shown in Fig.~7. The spatial distribution of HC$_5$N is not constrained and we considered two cases: a)~HC$_5$N arises from $g_3$, and the emission is so extended that the line brightness is well approximated by the line main-beam temperature; b)~HC$_5$N arises from $g_2$, with a typical size of $18\arcsec$. Case a) can be ruled out as the best fitting solution provides a rotational temperature
$T_{rot}= 87 \pm 26\K$, which is not consistent with the kinetic temperature of that component (20--$25\K$).
For case b), the emission of the detected transitions can be reasonably well described by one gas component with $T_{rot}$= $(51.3\pm 8.9)\K$ and column density N(HC$_5$N)= $(1.2\pm 0.4)\times 10^{12}\cmmd$. This is consistent with the gas kinetic temperature of component $g_2$ ($\approx 60\K$).
Based on our analysis of the excitation conditions we conclude that the emission of HC$_5$N is dominated by $g_2$ and
the abundance [HC$_5$N] is $\approx 1.2\times 10^{-9}$.
The best fitting solution to the line profiles is shown in Fig.~5. We note that the HC$_3$N column density is $\simeq 4.0\times 10^{12}\cmmd$ in the same range of excitation conditions, hence the relative abundance ratio HC$_3$N/HC$_5$N $\simeq 3.3$.

\subsection{HC$_7$N}
The non-detection of HC$_7$N could be due to a lower abundance (like in TMC1) and/or to less favourable excitation conditions than for the smaller cyanopolyynes. The lowest frequency transition in the ASAI band is $J$=64--63 72187.9088 MHz has an $E_{up}$= $112.6\K$. Between 80 and 114 GHz, the frequency interval in which HC$_5$N lines are detected, HC$_7$N lines have $E_{up}$ between 140 and $250\K$. These values are higher than the highest $E_{up}$
transition of HC$_5$N detected.  We note that a factor of 2-3 less in abundance would suffice to hamper any detection. In practice, we could only place an upper limit on the HC$_7$N abundance of $\approx 5\times 10^{-8}$.
Observations at lower frequency with the Green Bank Telescope or the VLA, would help confirming or not the presence of this molecule in the shock.

\section{The protostar L1157-mm}
In order to better understand the impact of the B1 shock on the ambient gas chemistry, we conducted a similar analysis towards the envelope of the protostar L1157-mm, which drives the outflow responsible for the
shock L1157-B1. We searched for and detected the emission of HC$_3$N transitions between $J$= 8--7 and  $J$= 28--27, as well as
of deuterated isotopologue DC$_3$N  between $J$= 9--8 and $J$= 18--17 (Figs.~A1-A3). We note that the latter molecule was not detected in the shock. We failed to detect the emission from the rare $^{13}$C isotopologues.
Four transitions of HC$_5$N were detected ($J$= 31--30, 32--31, 34--33 and 40--39). Unfortunately,  the number of detected HC$_5$N lines is too low and the flux uncertainties are too large to allow a detailed modelling of the emission. They will no longer be considered in what follows.

The emission was analyzed under the LTE hypothesis and following the same methodology as for the shock region.
In order to ease the reading, we present in this Section only our results. The data and the radiative transfer modelling are presented in the Appendix.

Linewidths are  much narrower towards L1157-mm, with $\Delta v \simeq$ 1.5--$2.5\kms$. Overall, we detected less transitions towards L1157-mm (see Table A1); in particular, we missed the emission from the high-$J$ levels detected towards the shock (see Figs.~A1-A3).

As can be seen in Fig.~A2, the HC$_3$N emission can be described by the sum of two physical components: first, a low-excitation component with $T_{rot}$= $(10.0\pm0.3)\K$ and N(HC$_3$N)= $(5.3\pm 0.6)\times 10^{12}\cmmd$, adopting a size of $20\arcsec$ for the envelope, based on submm continuum observations of the region by Chini et al. (2001), and second, a higher-excitation component with $T_{rot}$= $(31.3\pm 1.2)\K$ and N(HC$_3$N)= $(7.2\pm 1.0)\times 10^{12}\cmmd$, adopting a size of $5\arcsec$.
We estimated the total \htwo\ column density of the envelope from LTE modelling of flux of the $^{13}$CO $J$=1--0 line, adopting an excitation temperature of $10\K$ and a source size of $20\arcsec$, and a standard abundance ratio [$^{13}$CO]/[H$_2$]= $1.6\times 10^{-6}$.  From this simple analysis, we derive a relative abundance X(HC$_3$N)= $9\times 10^{-10}$ for the cold protostellar envelope. The uncertainty on the size of the warm gas component and the \htwo\ column density makes it difficult to estimate  molecular abundances in that region. Interferometer observations are required to better estimate these abundances.

The DC$_3$N emission arises from transitions with $E_{up}$ in the range $18\K$--$70\K$ (Table A.1).
Analysis of the DC$_3$N lines  yields $T_{rot}$= $(22.9\pm1.9)\K$ and N(DC$_3$N)= $(3.4\pm 0.6)\times 10^{11}\cmmd$, adopting a size of $20\arcsec$ (Fig.~A4).
In the $E_{up} < 40\K$ range, the emission of the main  HC$_3$N isotopologue is dominated by the low-excitation component with $T_{rot}$= $10\K$ (Fig.~A2). Under the reasonable assumption that DC$_3$N is dominated by the same gas component, we obtain a relatively high value of the deuterium fractionation ratio D/H= 0.06. In addition, considering only the transitions of lowest $E_{up}$ between $18\K$ and $27\K$, a fit to the DC$_3$N data yields $T_{rot}$= $(10.9\pm 1.7)\K$, in even better agreement with the results for the low-excitation HC$_3$N component, and N(DC$_3$N)= $(5.2\pm 1.8)\times 10^{11}\cmmd$. This  supports
the reasonable assumption that DC$_3$N and HC$_3$N trace the same gas region and the corresponding D/H ratio is 0.10.

\section{Discussion}

\begin{table}
\centering{
\caption{Properties of the HC3N isotopologues towards L1157-B1 and the cold envelope of L1157-mm. Cases where no estimate could be obtained are indicated by "-".}
\begin{tabular}{llcccc}\hline
Species  &      & \multicolumn{2}{c}{L1157-B1} & \multicolumn{2}{c}{L1157-mm}   \\
         &                  & $g_3$       &    $g_2$       &              \\ \hline
HC$_3$N  & Size ($\arcsec$) & Extended    &    18          &  20          &   5     \\
         & N($\cmmd$)       & 2.5(13)     & 4.0(12)        &  5.3(12)     & 7.2(12)  \\
         & T($\K$)          & 20          & 60             &  10          & 30  \\
         & n($\htwo$)       & 2(6)        & 4(6)           &  -          &  - \\
DC$_3$N  & Size ($\arcsec$) & Extended    & -            &  20          &   -   \\
         & N($\cmmd$)       & < 5(10)     & -             & (3.4-5.2)(11)&  - \\
         & T($\K$)          & 25          & -             & 11-23        &  -\\ \hline
\end{tabular}
}
\end{table}

\begin{table}
\centering{
\caption{Cyanopolyyne abundances determined towards L1157-B1 and the cold envelope of L1157-mm. Cases where no estimate could be obtained are indicated by "-".}
\begin{tabular}{cccc}\hline
Species  &  \multicolumn{2}{c}{L1157-B1}    &   L1157-mm   \\
         &   $g_3$       &    $g_2$         &              \\ \hline
HC$_3$N  &    2.5(-8)    &    4.0(-9)       &   0.9(-9)    \\
HC$_5$N  &     -        &    1.2(-9)       &     -       \\ \hline
\end{tabular}
}
\end{table}

\subsection{Shock initial conditions}

We can obtain a qualitative picture of the shock impact on cyanopolyyne chemistry from the assumption that the chemical and excitation conditions in the pre-shock molecular gas are rather similar to those in the envelope of L1157-mm.
In the outer regions of the latter, we estimate a typical abundance of $\sim 10^{-9}$, which implies that the  molecular abundance of HC$_3$N has been increased by a factor of about 30 through the passage of the shock.

The deuterium fractionation of HC$_3$N is rather large in the protostellar envelope (D/H= 0.06), whereas DC$_3$N is not detected in the shocked gas of L1157-B1, with an upper limit of 0.002 on the D/H ratio in L1157-B1. We note that if the shock production of HC$_3$N is not accompanied by any DC$_3$N gas phase enrichment, then the deuterium fractionation ratio is simply 30 times less, and D/H  becomes 0.002, the upper limit obtained with ASAI. Hence, DC$_3$N was probably not affected by the same abundance enhancement as its main isotopologue. This implies a low deuterium fractionation for HC$_3$N produced in the shock, either from grain sputtering or gas phase reactions. In both cases, HC$_3$N must have formed in warm (hot) gas. This is also consistent with the absence of $^{13}$C fractionation of HC$_3$N.

\subsection{The HC$_3$N/HC$_5$N ratio}
The high sensitivity of the ASAI data has allowed us to detect the emission of HC$_5$N, for the first time in a shock.
Jaber Al-Edhari et al. (2017) have  studied the HC$_3$N and HC$_5$N abundance distribution in different Galactic environments.
The abundance values we found in component $g_2$ for HC$_3$N and HC$_5$N,  $4.0\times 10^{-9}$ and $1.2\times 10^{-9}$, respectively, are among the highest values reported towards Galactic objects. They are typically higher than those found towards hot corino and WCCC sources by one to two orders of magnitude.
In particular, we found that the HC$_5$N abundance in L1157-B1 is higher than in the hot corino of IRAS16293-2422 by a factor of about 30.

In their study, Jaber Al-Edhari et al. (2017) showed that a low ($\approx 1$) relative abundance ratio HC$_3$N/HC$_5$N is typical of "cold", low-luminosity objects like early protostars (first hydrostatic core candidates) and WCCC sources (see e.g. Sakai et al. 2008). Analysis of the molecular content of the L1157-mm envelope recently led Lefloch et al. (2017) to classify it  as a WCCC protostar. Then, a low abundance ratio X(HC$_3$N)/X(HC$_5$N) is to be expected in the L1157-mm envelope.
The rather low abundance ratio X(HC$_3$N)/X(HC$_5$N) measured in the shock component $g_2$ ($\approx 3.3$) suggests that the abundance of HC$_5$N has increased by a factor similar to that of HC$_3$N in the shock. In general, based on the models shown in Fontani et al. (2017), a low HC$_3$N/HC$_5$N abundance ratio points toward an early time chemistry where the injected new carbon (as CH$_4$, CO or another species) is rapidly used for carbon chain formation and only later  (re-)forms CO.
It would be interesting  to probe lower-excitation transitions of HC$_5$N to determine its abundance in the outer envelope of L1157-mm and the low-excitation component $g_3$ of the shock.

\subsection{Shock modelling}
\begin{figure}
\centering{
\includegraphics[width=0.9\columnwidth,keepaspectratio]{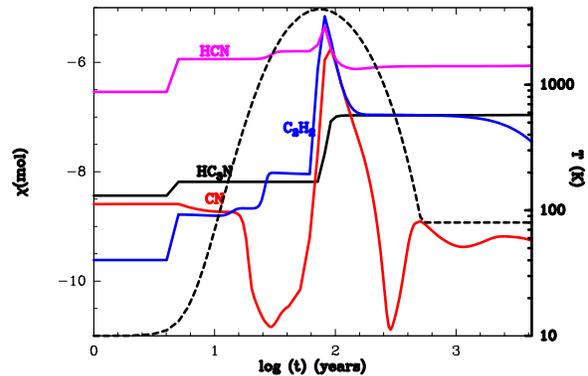}
\caption[]{Modelling of HC$_3$N (black) and its potential precursors HCN (magenta), C$_2$H$_2$ (blue) and CN (red), in L1157-B1. The temperature profile across the shock is displayed with the black dashed line.}
}
\end{figure}
In order to determine whether HC$_3$N arises in the shocked gas of the L1157-B1 cavity, we consider the best fit chemical and parametric shock model from previous works by our team
(Viti et al. 2011, Lefloch et al. 2016, Holdship et al. 2016), namely one where the pre-shock density is $10^5\cmmt$ and the shock velocity is $40\kms$. In Fig.~8, we plot the fractional abundance of HC$_3$N, and other selected species, as a function of time.
We find that the abundance of HC$_3$N increases at two different times: the first increase is due to mantle sputtering due to the passage of the shock. The second increase is in fact due to the reaction of CN with C$_2$H$_2$: CN + C$_2$H$_2$ $\rightarrow$ HC$_3$N + H. This channel has an increased efficiency when the temperature is much higher than $300\K$ due to the dependence of this reaction to temperature (Woon \& Herbst 1997) but, most importantly, due to the enhanced C$_2$H$_2$ and CN abundances when the temperature is at its maximum ($\sim 4000\K$). Indeed C$_2$H$_2$ and CN form via neutral-neutral reactions whose efficiency increases with temperature. In summary, the HC$_3$N abundance increases strongly as a consequence of the passage of the shock; this is consistent with the fact that observationally HC$_3$N appears to be enhanced (by a factor of 30) towards the shocked position of L1157-B1. Our modelling shows that the HC$_3$N abundance enrichment is dominated by the high-temperature gas phase reactions rather than sputtering of frozen HC$_3$N onto dust grains. This is consistent with the absence of differential fractionation between the three rare $^{13}$C isotopologues and the low deuterium enrichment when compared with the envelope of L1157-mm.

\section{Conclusions}
As part of ASAI, we have carried out a systematic search for cyanopolyynes HC$_{2n+1}$N towards the protostellar outflow shock region L1157-B1 and the protostar L1157-mm, at the origin of the outflow phenomenon. Towards L1157-B1, we confirm the detection of HC$_3$N lines $J$= 8--7 to $J$=29--28 and we report the detection of transitions from its $^{13}$C isotopologues $J$=9--8 to $J$=12--11. We have detected the HC$_5$N lines $J$=30--29 to $J$=43--42 for the first time in a shock. Towards L1157-mm, we detected the HC$_3$N lines $J$= 8--7 to $J$=28--27 and the deuterated isotopologue DC$_3$N lines $J$=9--8 to $J$=18--17. We summarize our main results as follows:

\begin{itemize}
\item~The abundance of HC$_3$N has been increased by a factor of about 30 in the passage of the shock.
\item~The upper limit we can place on the deuterium fraction ratio D/H in the shock (0.002) shows that the abundance enhancement of HC$_3$N was not accompanied by a significant abundance enhancement of DC$_3$N, which is consistent with a warm gas phase formation scenario. This is also supported by the observed lack of $^{13}$C fractionation of HC$_3$N.
\item~The rather low abundance ratio X(HC$_3$N)/X(HC$_5$N) measured in L1157-B1 suggests that HC$_5$N was also efficiently formed in the shock.
\item~A simple modelling based on the shock code of Viti et al. (2011) and the chemical code UCL\_CHEM,  adopting the best fit solution determined from previous work on L1157-B1 by our team, accounts for the above observational results. The HC$_3$N abundance increases in a first step as a result of grain mantle sputtering and, in a second step, as a result of the very efficient gas phase reaction of CN with C$_2$H$_2$.
\end{itemize}

\section*{Acknowledgements}
Based on observations carried out as part of  the Large Program ASAI (project number 012-12) with the IRAM 30m telescope.
IRAM is supported by INSU/CNRS (France), MPG (Germany) and IGN (Spain).  This work  was supported by the CNRS program "Physique et Chimie du Milieu Interstellaire" (PCMI) and by a grant from LabeX Osug@2020 (Investissements d'avenir - ANR10LABX56). E. M. acknowledges support from the Brazilian agency FAPESP (grant 2014/22095-6 and 2015/22254-0). I.J.-S. acknowledges the financial support received from the STFC through an Ernest Rutherford Fellowship (proposal number ST/L004801/1).

\appendix

\section{Cyanopolyyne emission in L1157-mm}
\begin{table*}
\centering{
  \caption{Spectroscopic  and observational parameters of the lines from  HC$_3$N and its isotopologues identified towards the protostar L1157-mm. We give in bracket the statistical uncertainties on the observational parameters, as derived from a gauss fit to the profiles.}
\begin{tabular}{lrrcccrrrr}
\hline
Transition  & Frequency & $E_u$ & $A_{ul}$ & HPBW & $\eta_{mb}$  & $\int T_{mb}dv$  & FWHM & $V_{lsr}$ \\
$J\rightarrow J-1$ & MHz &  K &  s$^{-1}$  & $^{\prime\prime}$ &  &   mK km s$^{-1}$ & km s$^{-1}$ & km s$^{-1}$ \\
\hline
HC$_3$N & & & & & & & &   \\
8 -- 7&	72783.822   & 15.7&	2.94(-5)& 33.8&	0.86&	338(4)& 2.3(0.8)&  3.0(0.1)	\\
9 -- 8&	81881.468   & 19.7&	4.21(-5)& 30.0&	0.85&	305(4)& 1.7(0.7)&  2.3(0.1)	\\
10 -- 9& 90979.023  & 24.0&	5.81(-5)& 27.0&	0.84&	330(4)& 1.7(0.6)&  2.9(0.1)	\\
11 -- 10& 100076.392& 28.8&	7.77(-5)& 24.6&	0.84&	264(7)& 1.5(0.6)&  2.8(0.1)	\\
12 -- 11& 109173.634& 34.1&	1.01(-4)& 22.5&	0.83&	267(7)& 1.4(0.5)&  2.4(0.1)	\\
14 -- 13& 127367.666& 45.9&	1.62(-4)& 19.3&	0.81&	189(8)& 1.3(0.5)&  2.5(0.1) \\
15 -- 14& 136464.411& 52.4&	1.99(-4)& 18.0&	0.81&	178(9)& 1.4(0.4)&  2.6(0.1) \\
16 -- 15& 145560.960& 59.4&	2.42(-4)& 16.9&	0.79&	141(9)& 1.1(0.4)&  2.5(0.1)	\\
17 -- 16& 154657.284& 66.8&	2.91(-4)& 15.9&	0.77&   113(19)&1.1(0.4)&  2.7(0.1)	\\
18 -- 17& 163753.389& 74.7&	3.46(-4)& 15.0&	0.77&    99(12)&1.4(0.4)&  2.5(0.1)\\
23 -- 22& 209230.234&120.5& 7.26(-4)& 11.8&	0.67&	 72(9)& 1.5(0.2)&  2.5(0.1)	\\
24 -- 23& 218324.723&131.0& 8.26(-4)& 11.3&	0.65&	 31(10)& 1.5(0.5)& 2.8(0.2)	\\
27 -- 26& 245606.320&165.0& 1.18(-3)& 10.0&	0.62&	 27(10)& 0.9(0.2)& 2.2(0.1)	\\
28 -- 27& 254699.500&177.3& 1.32(-3)&  9.7&	0.60&	 20(12)& 1.0(0.2)& 1.1(0.1)	\\
\hline
DC$_3$N &	        &	   &		 &      &	   &         &     &           \\
9 -- 8  & 75987.138 & 18.2 & 3.39(-5)& 32.4 & 0.86 & 26(6) & 2.7(1.3) & 1.8(0.5) \\
10 -- 9 & 84429.814 & 22.3 & 4.67(-5)& 29.8 & 0.85 & 29(6) & 2.0(0.7) & 2.5(0.3) \\
11 -- 10& 92872.375 & 26.7 & 6.24(-5)& 26.4 & 0.84 & 25(4) & 1.8(0.3) & 2.9(0.1) \\
13 -- 12&109757.133 & 36.9 & 1.04(-4)& 22.4 & 0.83 & 31(11)& 1.9(1.2) & 2.1(0.5) \\
18 -- 17&151966.372 & 69.3 & 2.78(-4)& 16.2 & 0.77 & 19(3) & 1.3(0.2) & 0.4(0.1)\\
\hline
HC$_5$N	&		    &	   &		 &		&	   &			&				&		 \\
31 -- 30& 82539.039 & 63.4 & 6.04(-5)& 29.8 & 0.85 & 36(8)&	1.5(0.5) & 3.5(0.2)	\\
32 -- 31& 85201.346 & 67.5 & 6.64(-5)& 28.9 & 0.85 & 27(5)& 2.0(0.7) & 2.5(0.3)	\\
34 -- 33& 90525.889 & 76.0 & 7.98(-5)& 27.2 & 0.85 & 14(2)& 1.7(0.4) & 2.4(0.2)	\\
40 -- 39&106498.910 &104.8 & 1.30(-4)& 23.1 & 0.84 & 13(2)&	1.3(0.1) & 2.5(0.1)	\\
\hline
\end{tabular}
}
\end{table*}

\begin{figure*}
\centering{
\caption{Montage of some HC$_3$N lines detected towards L1157-mm.}
\includegraphics[width=1.8\columnwidth,,keepaspectratio]{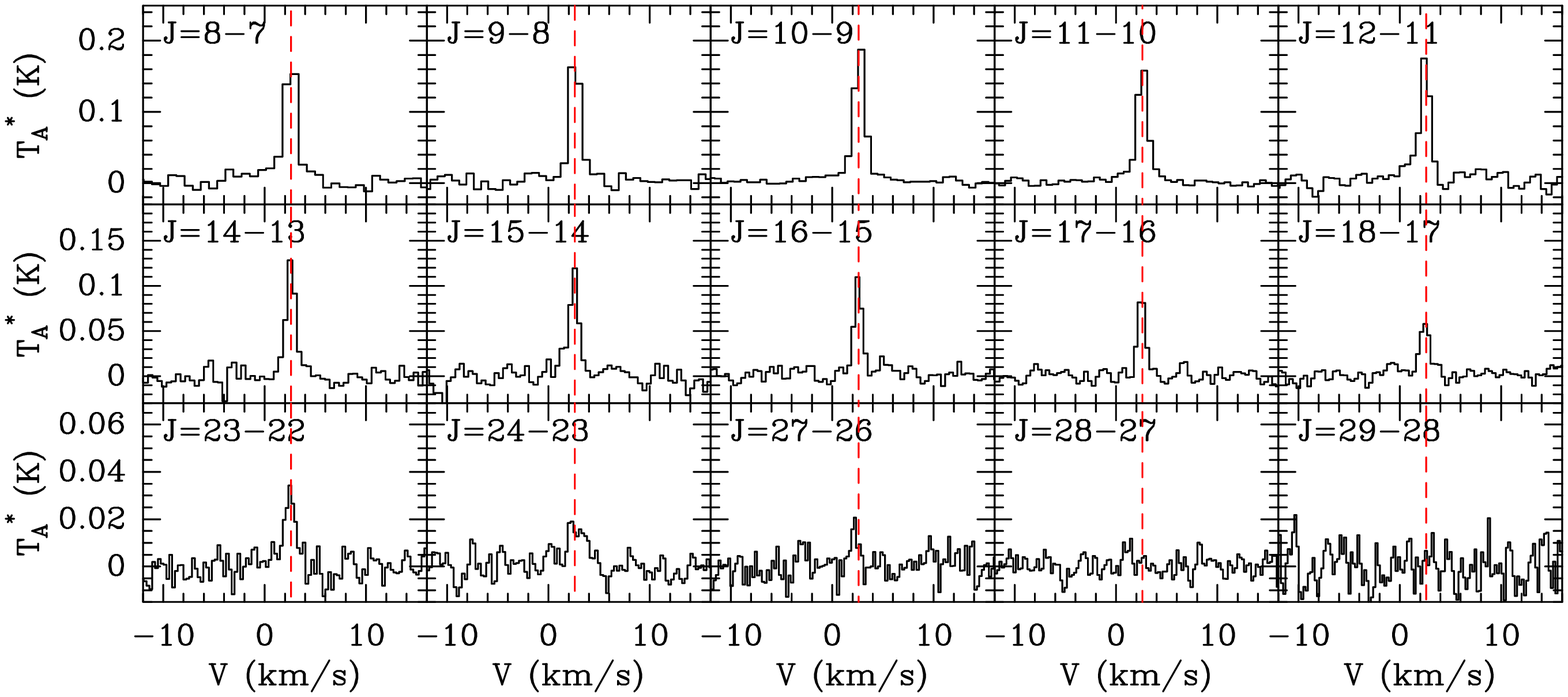}
}
\end{figure*}

\clearpage

\begin{figure}
\centering{
\caption{Rotational diagram analysis of HC$_3$N lines detected towards L1157-mm. The emission was corrected for the coupling between the source and the telescope beam assuming a typical size of $20\arcsec$ for the protostellar envelope. The fits to the two physical components accounting for the emission are shown in red and blue, respectively. In red,  the cold component at $T_{rot}$= $10.0\pm 0.3\K$ and N(HC$_3$N)= $(5.3\pm 0.6)\times 10^{12}\cmmd$; in blue, the warmer component at $T_{rot}$= $31.3\pm 1.2\K$ and N(HC$_3$N)= $(7.2\pm 1.0)\times 10^{12}\cmmd$. }
\includegraphics[width=\columnwidth,keepaspectratio]{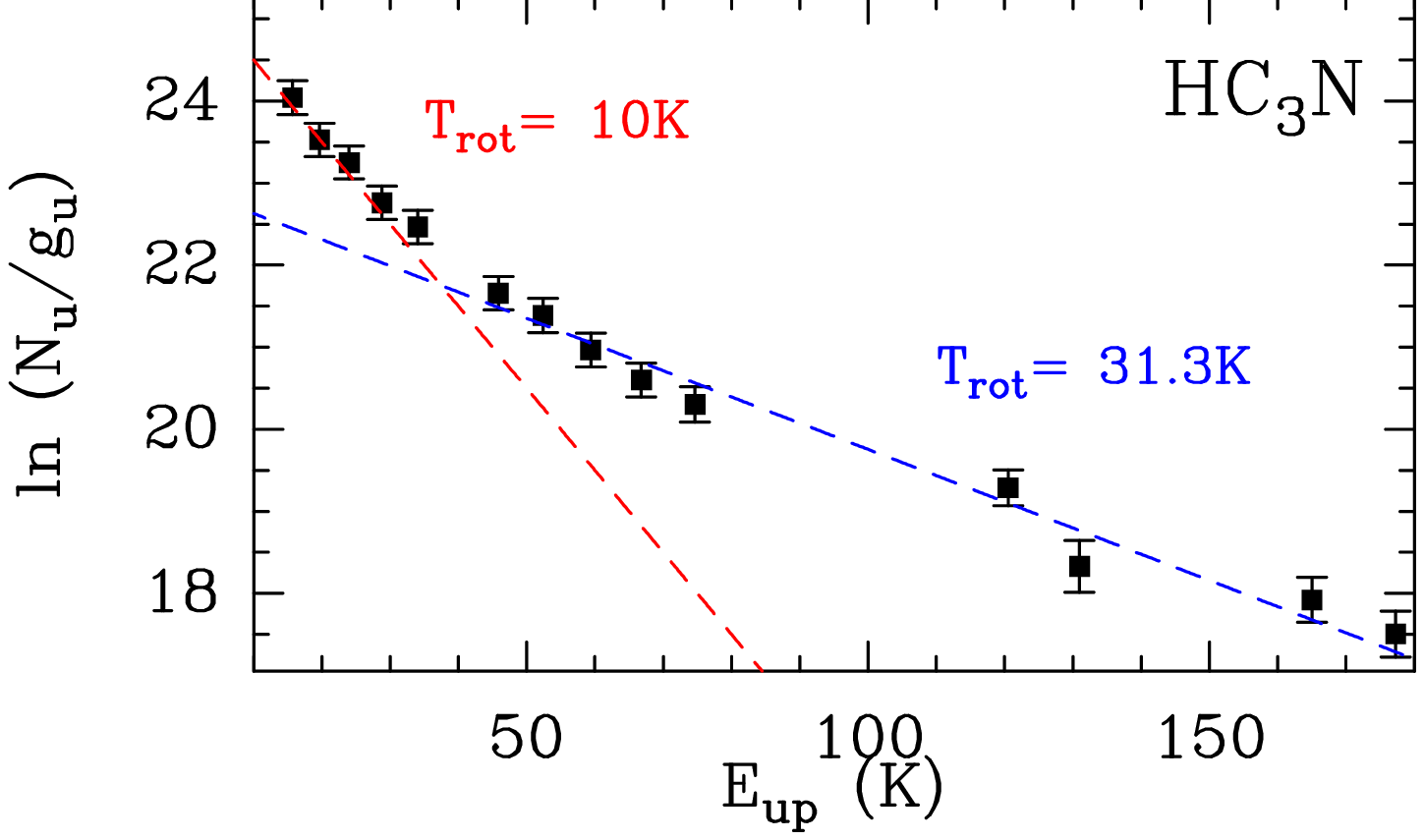}
}
\end{figure}

\begin{figure}
\centering{
\caption{Montage of some DC$_3$N lines detected towards L1157-mm. We have superposed in red the best LTE fit to the emission with $T_{rot}$= $22.5\pm 1.6\K$ and N(DC$_3$N)= $(3.4\pm 0.5)\times 10^{11}\cmmd$ and a source size of $20\arcsec$. }
\includegraphics[width=0.8\columnwidth]{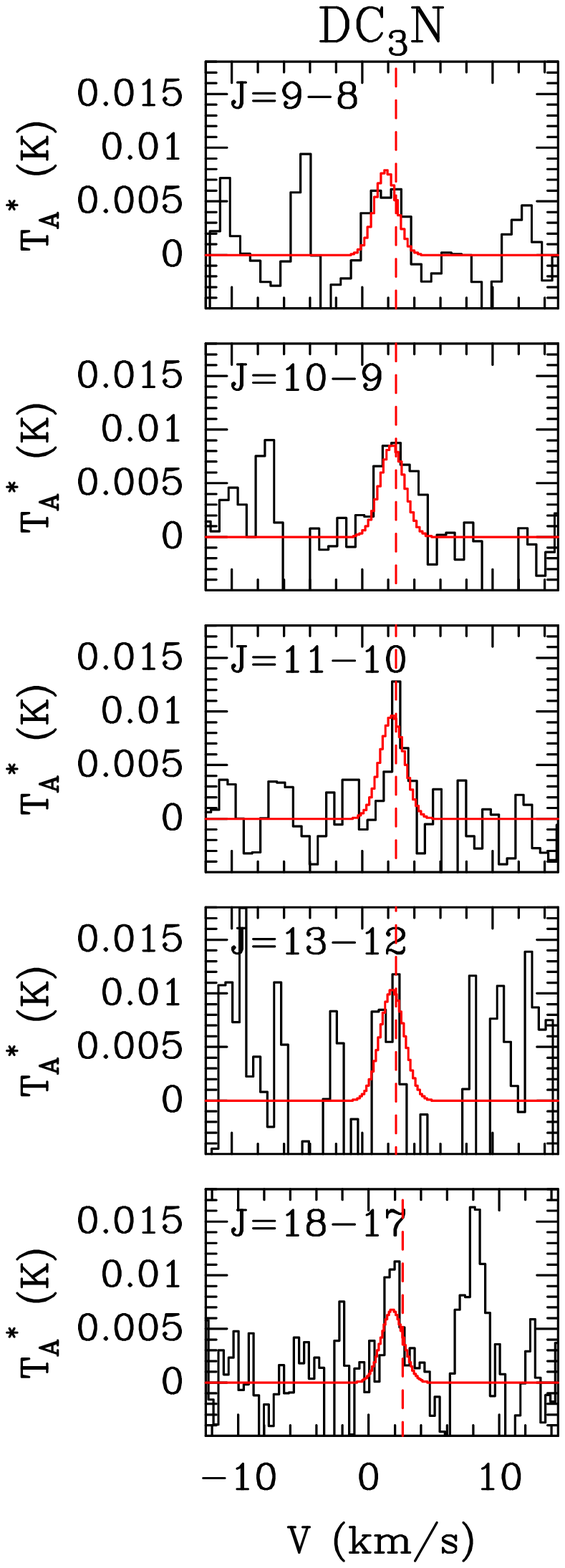}
}
\end{figure}

\begin{figure}
  \centering
  \includegraphics[width=\columnwidth,keepaspectratio]{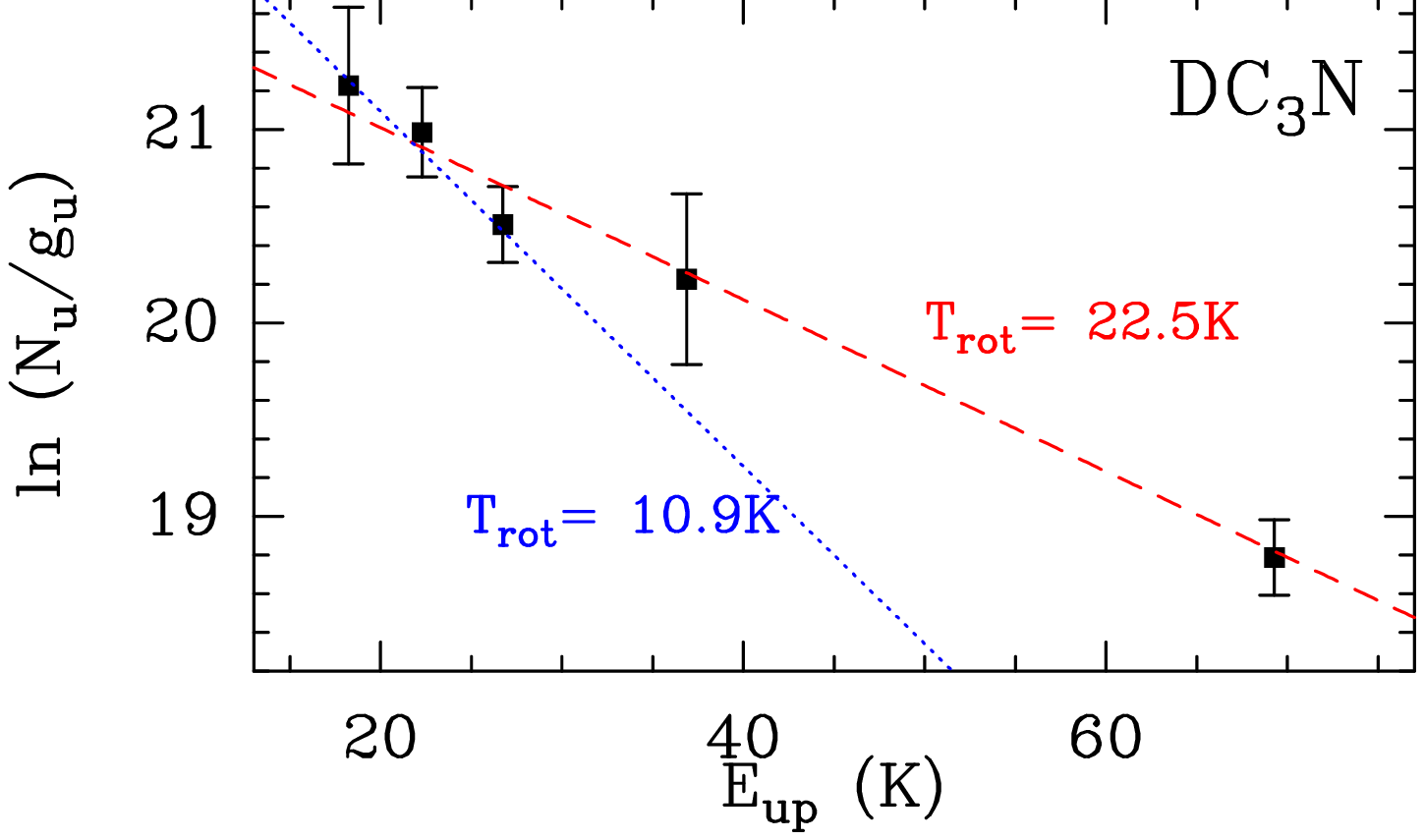}
  \caption{Rotational diagram analysis of DC$_3$N emission. The emission was corrected for the coupling between the source and the telescope beam assuming a typical size of $20\arcsec$. The best fit (dashed red line) is obtained for $T_{rot}$= $22.5\pm 1.6\K$ and N(DC$_3$N)= $(3.4\pm 0.5)\times 10^{11}\cmmd$. The fit obtained when considering only the data in the $E_{up} < 40\K$ range yields $T_{rot}$= $10.9\pm 1.7\K$ and N(DC$_3$N)= $(5.2\pm 1.8)\times 10^{11}\cmmd$; it is drawn by the blue dotted line. }
    \label{rotational}
\end{figure}



\end{document}